\documentclass[prd,aps]{revtex4}
\usepackage{graphicx,latexsym}
\input psfig
\pssilent
\begin{document}
\title{Consistent discretizations: the Gowdy spacetimes}
\author{Rodolfo Gambini$^{1}$, Marcelo Ponce$^{1}$ 
and Jorge Pullin$^{2}$}
\affiliation {
1. Instituto de F\'{\i}sica, Facultad de Ciencias,
Igu\'a 4225, esq. Mataojo, Montevideo, Uruguay. \\ 
2. Department of Physics and Astronomy, Louisiana State University, 
Baton Rouge, LA 70803-4001}
\date{May 9th 2005}

\begin{abstract}
  We apply the consistent discretization scheme to general relativity
  particularized to the Gowdy space-times. This is the first time the
  framework has been applied in detail in a non-linear
  generally-covariant gravitational situation with local degrees of
  freedom. We show that the scheme can be correctly used to
  numerically evolve the space-times.  We show that the resulting
  numerical schemes are convergent and preserve approximately the
  constraints as expected.
\end{abstract}

\maketitle

\section{Introduction}

The ``consistent discretization'' approach \cite{DiGaPu} has proven
quite attractive to deal with conceptual issues in quantum gravity
\cite{DiGaPubrasil}. In this approach one approximates general
relativity with a discrete theory that is constraint-free. As a
consequence, it is free of the hard conceptual issues that plague
quantum gravity. One can, for instance, solve the ``problem of time''
\cite{greece} and one discovers that there is a fundamental
decoherence \cite{njp,cqg} induced in quantum states that can yield
the black hole information puzzle unobservable \cite{prl,piombino}.

A looming question in this approach is how well can the
constraint-free discrete theory approximate general relativity?  In
this approach one starts with a Lagrangian, discretizes space-time and
then works out discrete equations of motion that determine not just
the usual variables of general relativity but the lapse and the shift
as well. This has the attractive aspect of yielding a consistent set
of discrete equations (they can all be solved simultaneously, unlike
in usual discretizations where the constraints are not preserved upon
evolution; it should be emphasized that preserving the constraints is
becoming a central focus in state-of-the-art numerical relativity.)
However, the resulting equations for determining the lapse and the
shift are non-polynomial and of a rather high degree. Since these
equations have no counterpart in the continuum theory, one does not
have an underlying mathematical theory to analyze them as one has in
the case of the other discrete equations. For instance, it is not
obvious that the solutions of the equations will be real. Or that they
do not oscillate wildly upon each step of evolution. We had probed
these issues in some detail in some cosmological models
\cite{cosmo}. There the consistent discretization approach works well,
approximating the continuum theory in a controlled fashion and
covering all of the phase space. But in the cosmological case the
equations also simplify significantly.  Moreover, the diffeomorphism
constraint does not play a role. In addition to this, it is known that
discretizations of evolution equations can pose unique challenges.
For instance, it is not clear that the resulting scheme produced by
the consistent discretizations is hyperbolic in any given sense. With
regular discretizations, the use of formulations that are not
hyperbolic has led to problems.  The type of discretizations that
arise in the consistent discretization approach are also ``forward in
time centered in space''. This is due to the fact that the use of
centered derivatives in time complicates the construction of the
canonical formulation that is central to the use of consistent
discretizations in quantum gravity. All this raised significant
suspiciousness about how well these discretizations could approximate
the continuum theory.  An encouraging point was that at least for
linearized gravity, the consistent discretizations yield a ``mimetic''
discretization \cite{mimetic} that appears to be stable. But the
linearized case has many peculiarities that do not carry over to the
non-linear case (namely the mimetism). It is also true that at the
classical level the ``consistent discretization'' approach may be
considered as an extension of the idea of variational integrators to
singular Lagrangians (systems with constraints treated in the Dirac
fashion).  Variational integrators have proven to have many desirable
properties, at least for unconstrained systems (see \cite{variational}
for a recent review), but they have only been applied in a limited way
to systems with constraints (only holonomic or at best constraints
linear in the momenta \cite{mclc}) without treating them in the usual
Dirac fashion.

It is clear that the viability of the approach has to be probed in
situations with field-theoretical degrees of freedom. A situation of
this type arises in the Gowdy cosmologies \cite{gowdy}. These are
spatially compact space-times with two space-like commuting Killing
vector fields. This allows the introduction of coordinates where the
metric depends only on time and an angular variable which we will call
$\theta$. The consistent discretization approach applied to this model
yields quite involved equations, as we will see. They cannot be solved
analytically and have to be tackled numerically. Moreover, since we
are dealing with non-linear coupled algebraic systems, the solution
has to be constructed approximately, and the resulting systems are
rather large. In addition to this, the presence of global constraints
derived from the compact topology yields some of the systems
unsolvable directly and this issue has to be addressed again by an
approximate technique.

It should be made clear from the outset that we are not at all
interested in being competitive in the well developed field of
numerical simulations of Gowdy cosmologies (see Berger \cite{berger}
for a review). Usually numerical simulations of Gowdy take
advantages of special gauges and other features that we will not
take into account in our approach. 

To simplify matters we will concentrate on the ``single polarization''
Gowdy case and we will further restrict to a particular subcase that
can be seen as a ``minisuperspace'' by setting one of the
variables and its conjugate momentum to zero.  From the point of view
of the evolution equations and the dynamics in terms of coordinate
dependent quantities, the complexity of the equations is virtually
similar to the full single-polarized case, one just is faced with a
smaller number of equations. Moreover, if one were to write the full
one polarization case and take suitably restricted initial data, the
evolution is preserved exactly within the minisuperspace
considered.

The organization of this article is as follows. In the next section
we briefly review the formulation of Gowdy cosmologies mostly to
fix notation. We then proceed to work out the consistent
discretization. In section III we discuss the numerical scheme
and present the numerical results.

\section{Gowdy cosmologies}
\subsection{Continuum formulation}
Gowdy cosmologies are space-times with two space-like commuting
Killing vector fields. We take the topology of the spatial slices to
be a three-torus $T^3$.  Following Misner \cite{misner}, we
parameterize the spatial metric by,
\begin{equation}
ds^2 = e^{-\tau -(\lambda/2)} d\theta^2+e^{2\tau} \left(e^\beta
d\sigma^2+e^{-\beta} d\delta^2\right)
\end{equation}
where functions $\tau,\lambda,\beta$ depend 
on time $t$ and on one of the spatial variables which we call 
$\theta \in [0,2\pi]$. The other two spatial variables are $\sigma$
and $\delta$ and will play no role from now on. The Hilbert
action particularized for these kinds of metrics reads,
\begin{equation}
S = {1 \over 2\pi} \int dt \int d\theta \left(
p_\lambda \dot{\lambda} + p_\tau \dot{\tau} +p_\beta \dot{\beta}
-{\bar N}_\mu {\bar C}^\mu,\right)\label{action}
\end{equation}
where ${\bar N}_\mu, \mu=0,1$ are related to 
the lapse and the single remaining 
component of the shift (Lagrange multipliers)
by ${\bar N}=N g^{-1/2}=N\exp(-3\tau/2+\lambda/4)$,
${\bar N}_\theta=N_\theta \exp\left(\tau+\lambda/2\right)$.
Also ${\bar C}^\mu$ are related to the
Hamiltonian and the single remaining component of the momentum
constraint by the obvious rescalings,
\begin{eqnarray}
{\bar C}^0&=&{1 \over 2} p_\beta^2 +p_\lambda p_\tau+ {1 \over 2} 
e^{4 \tau}
\left(\beta'\right)^2
+e^{4\tau}\left(4\tau''+8\left(\tau'\right)^2+\tau'\lambda'\right),\\
{\bar C}^\theta &=& 4 p_\lambda'+p_\tau \tau'
+p_\beta \beta'+p_\lambda \lambda'.
\end{eqnarray}
We have also chosen the irrelevant spatial coordinates in such a
way that $\int d\sigma \int d\delta=8$.

The theory can be consistently reduced by setting $\beta=p_\beta=0$.
This corresponds to a one-parameter family of space-times. Since in
our approach the coordinates are not fixed, the treatment is highly
non-trivial, and the problem appears to be ``infinite dimensional''
(although one is strictly speaking dealing with a zero dimensional
situation, as is the case in minisuperspaces). The exact solution of
the evolution equations for this case can be written, in a given gauge
choice (as shown by Misner). If one chooses lapse equal one, shift
equal zero, $\tau'=0$, $(p_\lambda)'=0$ one completely fixes gauge and
the solution results $\tau=c t$, $p_\lambda=c$, $p_\tau=0$ and
$\lambda=0$. We will see that in our approach we can choose initial
data where the shift approximately vanishes upon evolution. This is
good since it will allow us to compare scalars computed in the
numerical solution  with those
of the exact solution. If the shift is non-vanishing then the
situation complicates since one does not know at which points to
compare the invariants (one is facing the classic ``metric equivalence
problem'', i.e., comparing metrics in two different coordinate systems
to see if they are the same).

\subsection{Consistent discretization}

We now proceed to work out the consistent discretization of the
theory. We assume that the time coordinate is related to a discrete
variable $n$ by $t=n\Delta t$, $n=0\ldots N$. The spatial coordinate
$\theta$ is also discretized $\theta=m \Delta \theta$ with $m=0\ldots
M$. The action then becomes,
\begin{eqnarray}
S&=&\sum_{n=0} L(n,n+1)\nonumber\\
&=&\sum_{n=0}^N \sum_{m=0}^M \left\{\begin{array}{c}\\\\\end{array}
p_\lambda(n,m) \left(\lambda(n+1,m)-\lambda(n,m)\right)
+
p_\tau(n,m) \left(\tau(n+1,m)-\tau(n,m)\right)\right.\\
&&-M(n,m)\left[p_\lambda(n,m) p_\tau(n,m)
+e^{4\tau(n,m)}
\left(4\tau(n,m+1)+4\tau(n,m-1)-8\tau(n,m)\begin{array}{c}\\\\\end{array}\right.\right.\nonumber\\
&&\left.\left.+8\left(\tau(n,m+1)-\tau(n,m-1)\right)^2
+\left(\tau(n,m+1)-\tau(n,m-1)\right)
\left(\lambda(n,m+1)-\lambda(n,m-1)\right)\right)\right]\nonumber\\
&&-N(n,m)\left[\begin{array}{c}\\\\\end{array}
4p_\lambda(n,m+1)-4 p_\lambda(n,m-1)\right.\nonumber\\
&&\left.\left.+p_\lambda(n,m)\left(\lambda(n,m+1)-\lambda(n,m-1)\right)
+p_\tau(n,m)\left(\tau(n,m+1)-\tau(n,m-1)\right)  
\begin{array}{c}\\\\\end{array}
\right]
\right\}
\nonumber
\end{eqnarray}
where we have rescaled the momenta $p(n,m)\equiv p(t,\theta)\Delta
\theta$, and rescaled and relabeled the shift
$N(n,m)\equiv\bar{N}_\theta(t,\theta)\Delta t /(\Delta \theta)$, and
the lapse $M(n,m)\equiv\bar{N}(t,\theta)\Delta t /(\Delta
\theta)$. The reader should keep in mind that in the rest of this
paper when we refer to the ``lapse'' we are really referring at the
lapse times the time interval in the discretization divided by the
spatial interval.  For example, we will therefore encounter statements
like ``making the lapse small'' signifying that the discretization
step, measured in an invariant fashion, is becoming smaller.

We now proceed to define the canonical variables from the 
action. The reader should not be confused by the use of variables
named ``p'' in the action; in the consistent discretization
scheme all variables are initially treated as configuration variables (i.e.
one is working in a first order formulation of the theory, see
\cite{jmp} for a full discussion.)
One defines canonical momenta at instant $n$ and at instant
$n+1$. Let us consider for instance the variable $\lambda$. 
We define the canonical momenta at instant $n+1$ by,
\begin{equation}
P^\lambda(n+1,m)\equiv {\partial L \over \partial \lambda(n+1,m)}=
p_\lambda(n,m)
\end{equation}
and at instant $n+1$, using the Lagrange equations of motion one
has,
\begin{eqnarray}
P^\lambda(n,m)\equiv -{\partial L \over \partial \lambda(n,m)}&=&
p_\lambda(n,m)
+M(n,m-1)\left(\tau(n,m)-\tau(n,m-2)\right)
-M(n,m+1)\left(\tau(n,m+2)-\tau(n,m)\right)\nonumber\\
&&+N(n,m-1)p_\lambda(n,m-1)+N(n,m+1)p_\lambda(n,m+1).
\end{eqnarray}

One can combine these equations to eliminate the variable
$p_\lambda(n,m)$ from the model and yield an evolution equation
in terms of variables that are genuinely canonically conjugate,
\begin{eqnarray}
P^\lambda(n+1,m)&=&P^\lambda(n,m)
-M(n,m) e^{4\tau(n,m)}\left(\tau(n,m)-\tau(n,m+1)\right)
-M(n,m-1) e^{4\tau(n,m-1)}\left(\tau(n,m)-\tau(n,m-1)\right)
\nonumber\\
&&+N(n,m) P^\lambda(n+1,m)
-N(n,m-1) P^\lambda(n+1,m-1).\label{1}
\end{eqnarray}  

A similar procedure yields an (implicit) evolution equation for
$P^\tau$,
\begin{eqnarray}
P^\tau(n+1,m)&=&P^\tau(n,m)\nonumber\\
&&-
M(n,m)\left(
4 e^{4\tau(n,m)}
\left(4\tau(n,m+1)+4\tau(n,m-1)-8\tau(n,m)
+8\left(\tau(n,m+1)-\tau(n,m)\right)^2\nonumber\right.\right.\\
&&+\left.\left.\left(\tau(n,m+1)-\tau(n,m)\right)
\left(\lambda(n,m+1)-\lambda(n,m)\right)\frac{}{}\right)
\right.\nonumber\\
&&\left.+
4 e^{4\tau(n,m)}  
\left(-8-16\tau(n,m+1)+16\tau(n,m)+\lambda(n,m)-\lambda(n,m+1)\right)
\right)\nonumber\\
&&+N(n,m)P^\tau(n+1,m)-N(n,m-1)P^\tau(n+1,m-1)-4 M(n,m+1) 
e^{4\tau(n,m+1)}\nonumber\\
&&-M(n,m-1)e^{4\tau(n,m-1)}
\left(4+16\tau(n,m)-16\tau(n,m-1)+\lambda(n,m)-\lambda(n,m-1)\right),
\label{2}
\end{eqnarray}
and for $\lambda$,
\begin{equation}
\lambda(n+1,m)=\lambda(n,m)+M(n,m)P^\tau(n+1,m)+N(n,m)
\left(-4+\lambda(n,m+1)-\lambda(n,m)\right)+4 N(n,m-1)
\label{5}
\end{equation}
and $\tau$,
\begin{equation}
\tau(n+1,m)=\tau(n,m)+M(n,m)P^\lambda(n+1,m)+N(n,m)
\left(\tau(n,m+1)-\tau(n,m)\right).\label{6}
\end{equation}

A similar treatment for the Lagrange multipliers (defining their
canonical momenta and combining the equations at $n$ and $n+1$) yields
the ``pseudoconstraints'',
\begin{eqnarray}
&&4P^\lambda(n+1,m+1)-4P^\lambda(n+1,m)\nonumber\\
&&+P^\lambda(n+1,m)\left(\lambda(n,m+1)-\lambda(n,m)\right)+P^\tau(n+1,m)
\left(\tau(n,m+1)-\tau(n,m)\right)=0\label{3}\\
&&P^\lambda(n+1,m)P^\tau(n+1,m)+
e^{4\tau(n,m)}
\left[4\tau(n,m+1)+4\tau(n,m-1)-8\tau(n,m)\right.\nonumber\\
&&\left.+8\left(\tau(n,m+1)-\tau(n,m)\right)^2
+\left(\tau(n,m+1)-\tau(n,m)\right)
\left(\lambda(n,m+1)-\lambda(n,m)\right)\right]=0.\label{4}
\end{eqnarray}

The terminology ``pseudoconstraints'' reflects the fact that
these equations can be seen as discretizations of the constraint
equations of the continuum theory. However, they are not 
constraints for the discrete theory in the usual sense of the
word since they involve variables at different instants of time.

The set of equations (\ref{1}-\ref{4}) constitute six nonlinear
coupled algebraic equations for the six unknowns,
$\tau,\lambda,P^\tau,P^\lambda,N,M$, that is, they determine the
canonical variables and the Lagrange multipliers. In fact, this is an
oversimplified view of the situation, since the equations link
variables at different spatial points $m$. In order to solve them, one
has to take into account that all variables in the Gowdy problem are
periodic, for instance in a lattice with $mm$ points,
$\tau(0)=\tau(mm)$, etc. and then one is left with a system of
$6\times mm$ equations with $6\times mm$ unknowns. So if one decides
to use 10 spatial points these are 60 equations with 60 unknowns.

In fact, one does not need to tackle the system in an entirely coupled
fashion, since the two evolution equations for $\lambda$ and $\tau$
are explicit.  We will discuss the solution strategy in more detail in
the next section.

A final remark is that Van Putten proposed a scheme which also can be
characterized as determining the lapse and shift, but it differs
significant from ours (in particular it is not motivated
variationally). He also applied it to the Gowdy cosmologies \cite{vp}.

\section{Numerical results}
\subsection{Choice of initial data}
In order to evolve the equations we need to set initial data.
The choice of initial data is somewhat delicate, since it will
determine not only the solution but also the lapse and the
shift. Although one can give arbitrary initial data, it would
be desirable if the resulting lapse were of the same sign
across the spatial manifold. If one does not make this choice,
the scheme is able to evolve, but portions of the spacetime 
will be evolved forward in time and some portions it will 
evolve backward in time. There is nothing wrong with this,
but it is not what is traditionally considered in numerical
evolutions. We may also wish to choose a vanishing shift,
although this is not important. 

In order to ensure that the lapse and shift are approximately in line
with what we desire, we proceed in the following way.  We start with
some initial data for the variables $\tau(0),\lambda(0), N(0),M(0)$.
We use equations (\ref{3},\ref{4}) to determine
$P^\tau(1),P^\lambda(1)$ and then we use equations (\ref{1},\ref{2})
to determine $P^\tau(0),P^\lambda(0)$. This completes the initial data
set.  That is, we have given us $\tau,\lambda$ and the lapse and the
shift that we desire, and this determined the canonically conjugate
momenta.

At this point the reader may be confused. Usually in numerical
relativity one gives the metric and its canonical momenta at a given
hypersurface, and they are chosen in such a way that they satisfy the
constraints. To implement the consistent discretization evolution in
this case one would consider the metric as evaluated at $0$ and the
canonical momenta evaluated at $1$. One then chooses lapse and shift
at $0$ and evolves. Presumably one would choose a small lapse in order
to generate an evolution that is close to the continuum theory (recall
that we refer here to the rescaled lapse). One could do things the
other way around, choosing the metric at $1$ and the canonical
momentum at $0$ and evolve backwards. The choice we made to of initial
data is equivalent to finding a solution of the usual constraints at a
given hypersurface with $\lambda,\tau$ given.

\subsection{Evolution}

We need to give a prescription that, given
$\tau(n),\lambda(n),P^\tau(n),P^\lambda(n)$ determines the variables
at instance $n+1$ and in the process determines the Lagrange
multipliers. One could take the complete set of equations and solve
them simultaneously. We did not proceed like this. Since this was the
first exploration of the system we thought it would be more
instructive and offer more control on the situation to solve the
equations in sub-systems. We will see however, that in the end one
pays a price for this.

The scheme we used to solve the system is as follows: using equations
(\ref{3},\ref{4}) we determine $P^\lambda(n+1),P^\tau(n+1)$. We then
take these values together with
$\lambda(n),\tau(n),P^\tau(n),P^\lambda(n)$ and use the equations
(\ref{1},\ref{2}) to determine the Lagrange multipliers $N(n),M(n)$.
With these multipliers and the initial values, it is possible, using
(\ref{5},\ref{6}) to determine $\lambda(n+1),\tau(n+1)$, which would
complete the evolution process from level $n$ to $n+1$.

Unfortunately, there is a problem with this procedure. When we try to
use equations (\ref{1},\ref{2}) to determine $M(n),N(n)$ one finds
that the system is indeterminate. One can determine all the Lagrange
multipliers except one, which we can choose to be $M1\equiv M(n,m=1)$.
This has to do with a symmetry of the Gowdy problem. The action
(\ref{action}) only depends on derivatives of the variable $\lambda $,
therefore it is invariant under the addition of a constant to that
variable. Using Noether's theorem one can find out that there is a
conserved quantity, $\int P^\lambda(\theta) d\theta$. A similar result
holds in the discrete action, the conserved quantity becomes
$\sum_{m=0}^M P^\lambda(m)$.  If one imposes that this quantity be
conserved, this can be used to determine entirely the Lagrange
multipliers.  Another way of seeing this is to notice that one can
cast the conservation law as a constraint on the initial data,
\begin{equation}
\sum_{m=0}^M P^\lambda(n+1,m)[\lambda(n,m),\tau(n,m)] =
\sum_{m=0}^M P^\lambda(n,m)
\end{equation}
that is, writing the $P^\lambda$ at $n+1$ as a function of the
initial data and imposing the conservation one has a constraint
on the initial data. Imposition of this constraint is enough
to determine entirely the Lagrange multipliers.

This is a pathology of our choice of scheme of solution of the coupled
system of equations, dealing with it by parts. If we had chosen to
solve the entire non-linear system (\ref{1},\ref{6}) such system would
have been well defined and would have automatically preserved the
conserved quantity upon evolution. This would have, however, forced us
to deal with a significantly larger coupled non-linear system.

Numerically, the way we will implement the conservation is to choose
arbitrarily the value of one component of the lapse.  We then run the
evolution scheme we outlined and we will check if $\sum_{m=0}^M
P^\lambda$ has varied.  If it has, we will go back and adjust the
arbitrary component of the lapse until the conserved quantity is kept
constant. We will do this via a Newton--Raphson technique. The flow
diagram of the logic is shown in figure 1. For the solution of the
various non-linear systems we discussed we use the routine
TENSOLVE \cite{tensolve}.
\begin{figure}[h]
\centerline{\psfig{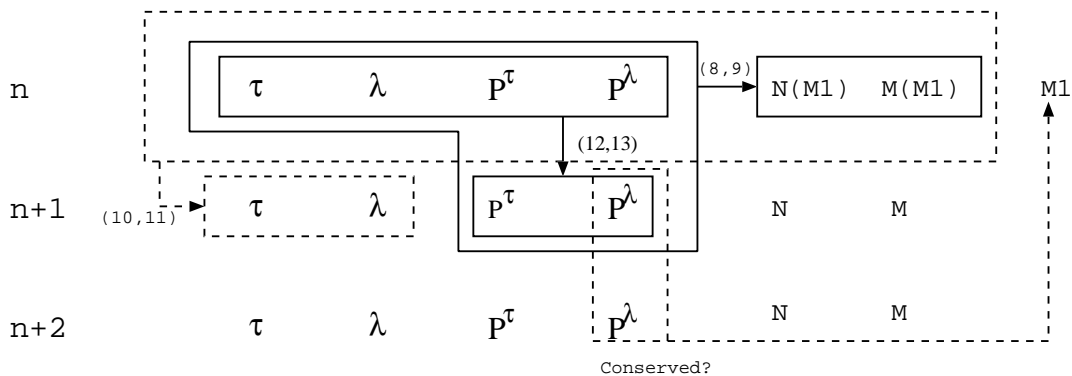}}
\caption{The logic of the evolution scheme.}
\end{figure}

\subsection{Results for the dynamical variables}

\begin{figure}[h]
\centerline{\psfig{file=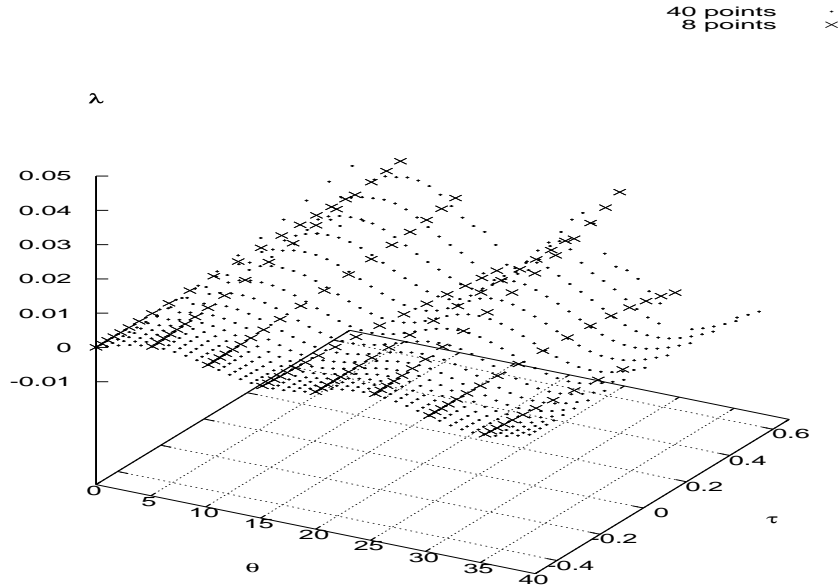,height=9cm,width=13cm}}
\caption{The variable $\lambda$ as a function of space ($\theta$)
and time, measured by the variable $\tau$. We show two resolutions,
one with 8 and another with 40 spatial points. The latter represents
what is practical to run on a workstation today. One sees that the
two resolutions track each other well for a while, but things
deteriorate as time increases (see for instance the third line of
crosses from the right). The run with 8 spatial points just does not
have enough resolution to track the features of the solution in
question, as can be seen from the figure.}
\label{figure1}
\end{figure}

In figure \ref{figure1} we show the variable $\lambda$ as a function
of the spatial points and as a function of time, for a simulation with
8 and 40 spatial points.  The time is measured by the variable
$\tau$, which is a good measure since it is invariant under coordinate
changes of the $t,\theta$ variables. Since we have chosen the shift to
be close to zero, and it is preserved that way upon evolution as we
shall see, the value of the angle $\theta$ can be taken to be an
invariant.  We see that the variable $\lambda$ exhibits the types of
oscillations of increasing frequency that one encounters in the exact
solution for the Gowdy spacetime.
The evolutions with different spatial
resolutions all have the same initial data for
$\lambda(0),\tau(0),M(0),N(0)$.  The explicit form of initial data
chosen are,
\begin{eqnarray}
\tau(0,i)&=&-0.5+0.01 \sin(2 \pi i/M),\\
\lambda(0,i) &=& 0.001\sin(2 \pi i/M)+ 0.0025 \sin(4 \pi i/M),\\
M(0,i)&=&0.0005,\\
N(0,i)&=&10^{-7} \sin(2 \pi i/M).
\end{eqnarray}

\begin{figure}[ht]
\centerline{\psfig{file=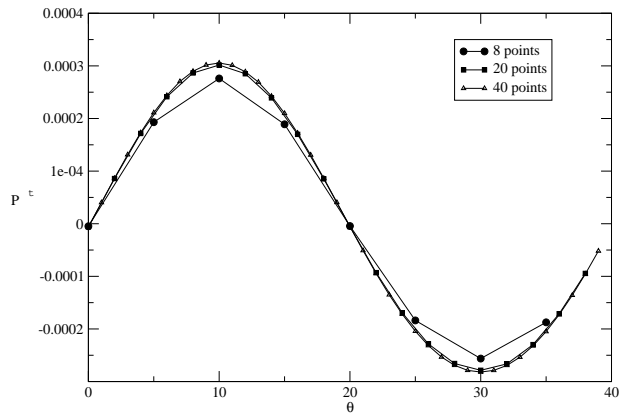,height=7cm,width=9cm}}
\caption{The variable $P^\tau$ in the initial data converges to a 
continuum solution of the pseudo-constraints, and would converge
to a solution of the continuum usual constraints of GR if the choice
of initial data for the lapse and shift were zero.}
\label{figure2}
\end{figure}

Following the initial data construction we outlined before, this
produces differing values of $P^\lambda(0),P^\tau(0)$ depending on the
number of points of the spatial grid. As shown in figure \ref{figure2}
there is convergence in the form of the resulting
$P^\lambda(0),P^\tau(0)$ when one refines the spatial grid.

The $\lambda(0),\tau(0),P^\lambda(0),P^\tau(0)$ produced by the 
initial data solving procedure, in the limit in which one makes
infinitely large the number of spatial points, is not a solution of
the initial value constraints of the continuum theory. However one
can approximate a solution of the constraints of the continuum
theory arbitrarily by choosing the value of the initial (rescaled)
lapse to be very small.

\begin{figure}[ht]
\centerline{\psfig{file=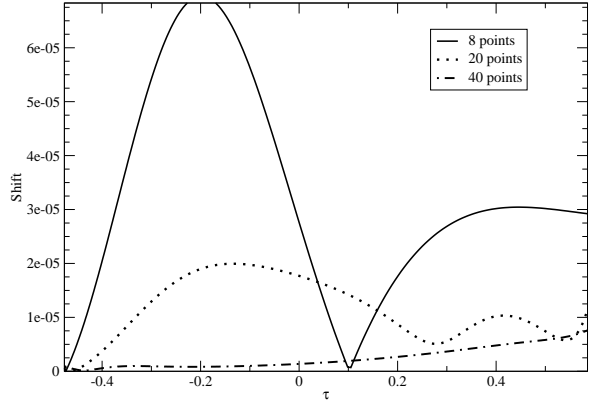,height=7cm,width=9cm}}
\caption{The L2 norm of the shift, which shows we are approximately in
a gauge with zero shift.}
\label{figure3}
\end{figure}

Figure \ref{figure3} shows the value of the average of the shift across
the grid as a function of time. We see that in all cases the shift is
small, and is smaller for higher resolutions. This is desirable in order
to compare with the exact solution which is in a gauge with zero shift.

\begin{figure}[ht]
\centerline{\psfig{file=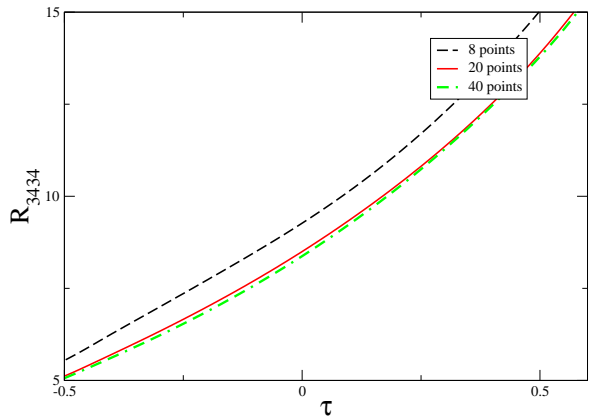,height=7cm,width=9cm}}
\caption{Convergence of the Riemann tensor for three different
resolutions. Since we are working with an almost vanishing shift, the
components of the Riemann tensor can be treated as observables and
compared among different resolutions directly. With a non-vanishing
dynamically determined shift this would not be possible. In such cases
one can only compare observables of the theory.}
\label{figure4}
\end{figure}

Figure \ref{figure4} shows the Riemann tensor for three
resolutions. More precisely, it represents component $R_{3434}$ (where
$3,4$ represent the ignorable coordinates) as a function of $\tau$ for
$\theta=\pi$. The reason we chose this component is that it behaves as
a scalar with respect to $t,\theta$ diffeomorphisms. Since we
determine the lapse and the shift dynamically, for different
resolutions we have different lapses and shifts and therefore
different coordinates.  As a consequence one cannot easily compare the
other components of the Riemann tensor for different resolutions. One
sees that the method converges, i.e. for smaller spatial separations
the value of the Riemann tensor approximates well that of flat
space-time.

\begin{figure}[ht]
\centerline{\psfig{file=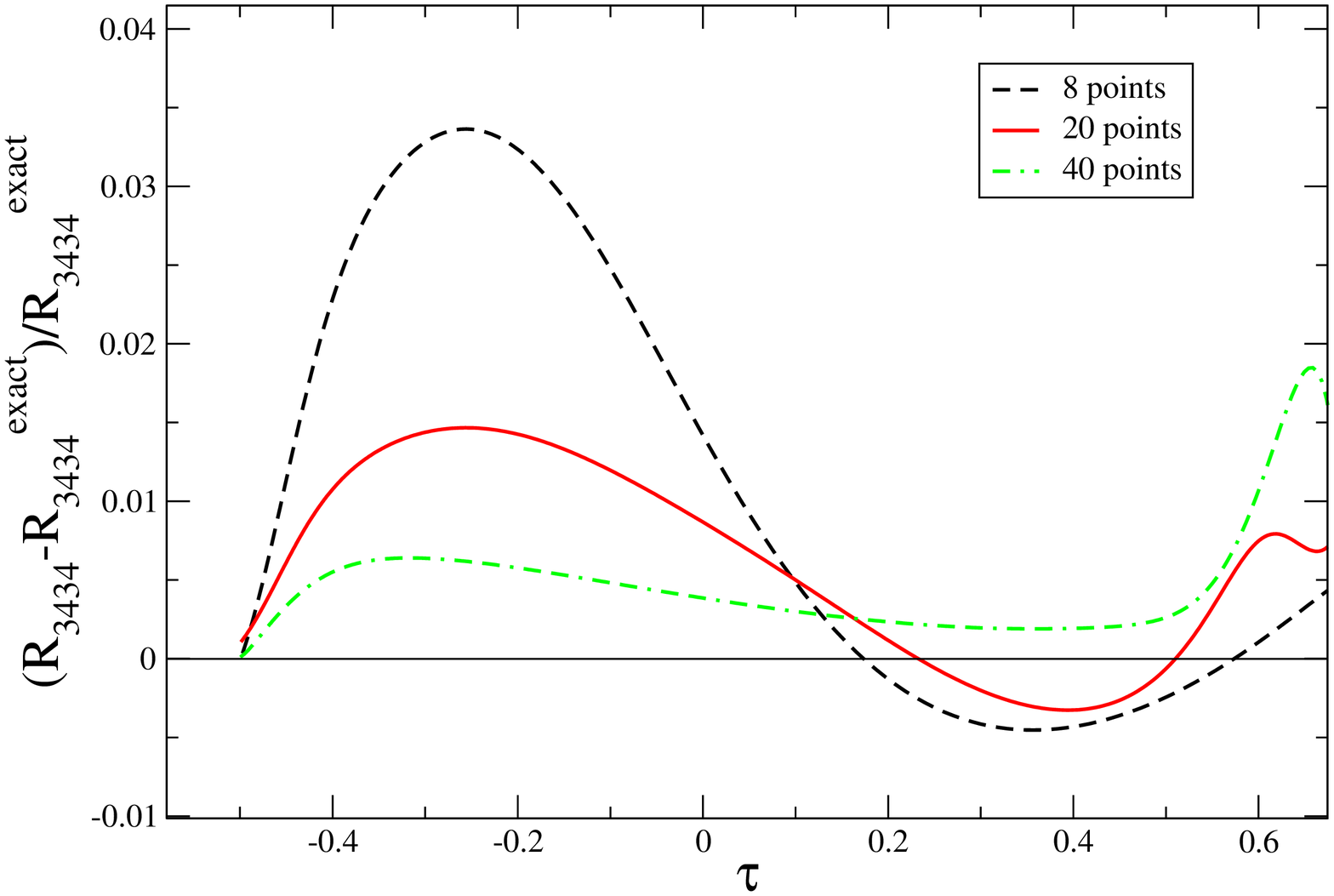,height=7cm,width=9cm}}
\caption{Relative error of the Riemann tensor compared to the
exact solution.}
\label{figure5}
\end{figure}

We know the exact form of the Riemann tensor. In figure \ref{figure5}
we use it to evaluate the relative error in the evaluation.  The exact
form is $R_{3434}=c^2 e^{-\tau}$ where $c$ is the only parameter
present in the minisuperspace (recall $\tau=ct$).  To avoid having to
determine the constant we actually plot in figure \ref{figure5} the
following quantity,
\begin{equation}
{{R_{3434}(\tau)\over R_{3434}(\tau(1))}-e^{-(\tau-\tau(1))}
\over e^{-(\tau-\tau(1))}}
\end{equation}
which yields the relative error.

\begin{figure}[ht]
\centerline{\psfig{file=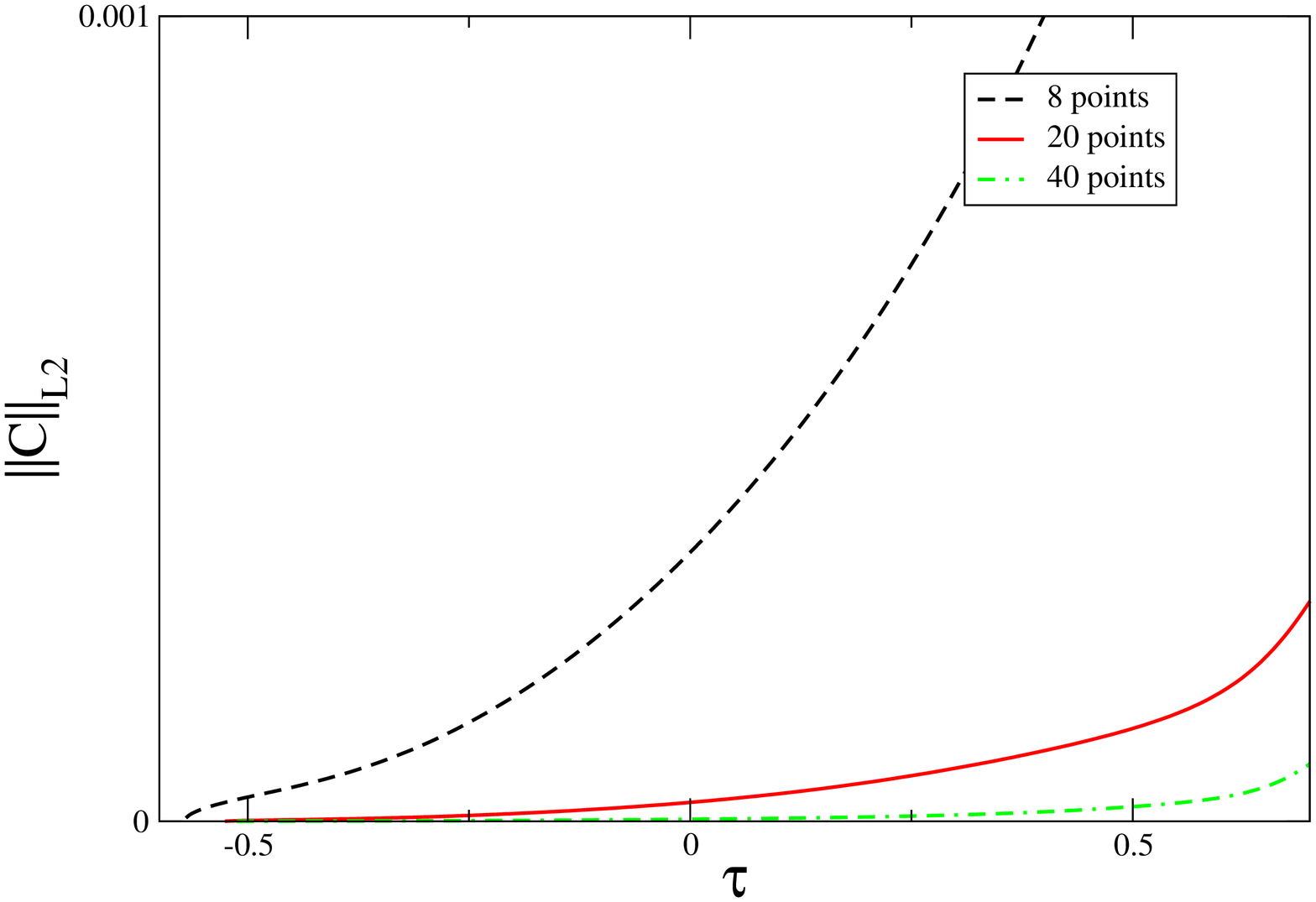,height=7cm,width=9cm}}
\caption{L2 norm of the constraints of the continuum theory
evaluated in the discrete theory. As can be seen they are 
preserved well and in a convergent fashion. As argued in the
text, the value of the constraints is a measure of the error
of the evolutions in the discrete theory and this plot 
confirms it.}
\label{figure6}
\end{figure}

\begin{figure}[ht]
\centerline{\psfig{file=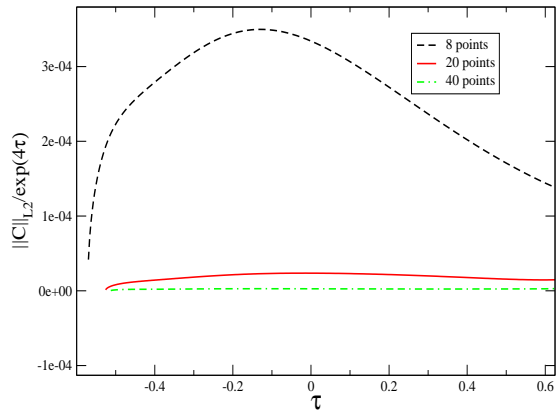,height=7cm,width=9cm}}
\caption{L2 norm of the constraints of the continuum theory
evaluated in the discrete theory normalized by the factor
$\exp(4\tau)$ that appears in the Hamiltonian constraint. 
One sees convergence with absence of growth in the 
normalized constraints. The reader may be puzzled as to why
the curves do not start at the same point. The reason is that
we are displaying the curves starting at the first iteration,
not at the initial data. For very low resolutions
the variables change quite a bit in the first iteration.}
\label{figure7}
\end{figure}

In figure \ref{figure6}
we show the L2 norm of the
square root of the sum of the squares of the constraints of the
continuum theory evaluated in the discrete theory, for three
resolutions. As expected, the magnitude of the constraints correlates
well with the error in the evolution scheme.
We are displaying the values of the constraints with any type of
normalization. This may be misleading. As we see, although there
is convergence there is also exponential growth. The growth can
be adscribed to the general exponential growth of variables in
Gowdy, particularly the factor $\exp(4\tau)$ that appears in the
Hamiltonian constraint. To compensate for this in figure \ref{figure7}
we display the same data divided by the offending factor, and one
cannot distinguish constraint growth.

\subsection{The flat sector}

If one chooses initial data for the evolution equations
(\ref{1},\ref{2},\ref{5},\ref{6}) such that $P^\lambda=0$ and
$\tau=0$, one can check that, in the continuum theory, one is
producing flat space-time. We would like to check if the discrete
theory is able to reproduce, at least in an approximate way,
this behavior. This is usually known as the ``gauge wave test''
as is described, for instance in the ``apples with apples'' 
(www.appleswithapples.org) project, and it is known to be a 
somewhat significant hurdle for codes to pass (at least in full 3D).

In the discrete theory the pseudo-constraints (\ref{3},\ref{4}) are
identically satisfied and the lapse and shift are free.  Let us start
by considering a slicing with $M=1,N=0$. There are only two evolution
equations left,
\begin{eqnarray}
P^\tau(n+1,m)&=&
P^\tau(n,m)+\lambda(n,m+1)-2\lambda(n,m)+\lambda(n,m-1)\\
\lambda(n+1,m)&=&P^\tau(n+1,m)+\lambda(n,m). 
\end{eqnarray}

These equations can be combined to yield a single equation for
$\lambda$, 
\begin{equation}
\lambda(n+1,m)=\lambda(n,m+1)+\lambda(n,m-1)-\lambda(n-1,m),
\end{equation}
and it is remarkable to notice that the exact solution of
the discrete equations for $\lambda$ is given by a plane wave,
\begin{eqnarray}
\lambda(n,m)=f(n+m)+g(n-m),
\end{eqnarray}
and therefore the space-time metric is manifestly a ``gauge
wave''. Also remarkable is that if one computes the discrete Riemann
tensor all its components vanish identically. In this sense, our
formalism aces one of the ``apples with apples'' tests exactly. If we
choose $M={\rm constant}$ instead of one we will get boosted slices
and again we will get the gauge wave as an exact solution and a
vanishing Riemann tensor. 
If we choose more complicated slices, then
the solution will be reproduced approximately. The resulting
equations are,
\begin{eqnarray}
P^\tau(n+1,m)-P^\tau(n,m)&=&M(n,m)(-8-\lambda(n,m+1)+\lambda(n,m))
-M(n,m-1)(4+\lambda(n,m)-\lambda(n,m-1))\nonumber\\
&&+N(n,m) P^\tau(n+1,m)
-N(n,m-1) P^\tau(n+1,m-1)-4 M(n,m+1)\\
\lambda(n+1,m)-\lambda(n,m)&=&M(n,m)
P^\tau(n+1,m)+N(n,m)(-4+\lambda(n,m+1)-\lambda(n,m)+4 N(n,m-1)
\end{eqnarray}

We have studied this case numerically and the code proves convergent
and long term stable (we could not find evolutions that crashed)
provided that one limits oneself to values of $M$ less than unity,
otherwise one can explicitly check that the evolution system has
eigenvalues of the amplification matrix larger than one (this is
just the Courant condition).

\section{Discussion}

We have applied for the first time the ``consistent discretization''
approach to a nonlinear situation with field theoretic degrees of
freedom, the Gowdy one polarization cosmologies. We see that one
can successfully evolve in a convergent and stable fashion for
about one crossing time. All of the expected features of
the approach are present. One encounters that the Lagrange 
multipliers are determined by the evolution and that at some
points they can become complex. Evolution can be continued by
backtracking and switching to a different root of the equations
that determine the Lagrange multipliers. We have not resorted
to this type of technique for the results presented up to now
in this paper. However, we have been able to extend the runs
to about ten crossing times ($\tau=2.8$) using these types
of techniques, but we have not carried out detailed 
convergence studies. The codes finally stop due to exponential
overflows in the resolution of the linear system derived from the
implicit nature of the equations.

The evaluation of the constraints of the continuum theory indicate
that they are well preserved in a convergent fashion and can be
used as independent error estimates of the solutions produced.

It should be noted that one has choices when one discretizes the
theory. We have chosen here to discretize the full model without
any gauge fixing. One could have, however, resorted to fixing 
gauge and discretizing a posteriori. There are various possibilities,
since one can also partially gauge fix and discretize. Partial
gauge fixings may be desirable in realistic numerical simulations
in order to have some control on the gauge in which the simulations
are performed. For instance, in binary black hole situations it has
been noted that the use of corrotating shifts is desirable. One
could envision a scheme in which one gauge fixes the shift to the
desired one and uses the consistent discretization to determine
the lapse.

The Gowdy example unfortunately has limitations as a test of the
scheme given the exponential nature of the metric components.
This prevents any scheme from achieving long term evolutions.
A scheme like ours that involves complex operations involving
many powers of the metric components is even more limited by
this problem. We have tried to improve things by using quadruple
precision in our Fortran code, but since one is battling 
exponential growth, this only offers limited help.

Summarizing, the scheme works, but given the limitations of the
example it is not yet clear that it will prove useful in long
term evolutions in other contexts. Turning to quantum issues,
the complexity exhibited in the determination of the lapse and
shift suggests that quantization of these models will have to
be tackled numerically. This paper can be seen as a first step
in this direction as well.

\section{Acknowledgments}
We wish to thank Luis Lehner and Manuel Tiglio for discussions
and Erik Schetter for comments on the manuscript.  This work was
supported by grant NSF-PHY0090091, NASA-NAG5-13430 and funds from the
Horace Hearne Jr. Laboratory for Theoretical Physics and CCT-LSU.

\end{document}